# Chirality-induced Magnet-free Spin Generation in a Semiconductor

Tianhan Liu[1,4,†,*], Yuwaraj Adhikari[1,†], Hailong Wang[2,†], Yiyang Jiang[3], Zhenqi Hua[1], Haoyang Liu[1], Pedro Schlottmann[1], Hanwei Gao[1], Paul S. Weiss[4,5], Binghai Yan[3], Jianhua Zhao[2,*], and Peng Xiong[1,*]

[1]Department of Physics, Florida State University, Tallahassee, Florida 32306, USA

[2]State Key Laboratory of Superlattices and Microstructures, Institute of Semiconductors, Chinese Academy of Sciences, Beijing 100083, China

[3]Department of Condensed Matter Physics, Weizmann Institute of Science, Rehovot 7610001, Israel

[4]Department of Chemistry and Biochemistry, University of California, Los Angeles, Los Angeles California 90095, USA

[5]California NanoSystems Institute and Departments of Bioengineering and Materials Science and Engineering, University of California, Los Angeles, Los Angeles California 90095, USA

[†]These authors contributed equally.

[*]Corresponding author emails: tianhanliu@g.ucla.edu, jhzhao@semi.ac.cn, pxiong@fsu.edu

**Abstract:**

Electrical generation and transduction of polarized electron spins in semiconductors are of central interest in spintronics and quantum information science. While spin generation in semiconductors has been frequently realized via electrical injection from a ferromagnet, there are significant advantages in nonmagnetic pathways of creating spin polarization. One such pathway exploits the interplay of electron spin with chirality in electronic structures or real space. Here, utilizing chirality-induced spin selectivity (CISS), we demonstrate efficient creation of spin accumulation in *n*-doped GaAs via electric current injection from a normal metal (Au) electrode through a self-assembled monolayer of chiral molecules (α-helix L-polyalanine, AHPA-L). The resulting spin polarization is detected as a Hanle effect in the *n*-GaAs, which is found to obey a distinct universal scaling with temperature and bias current consistent with chirality-induced spin accumulation. The experiment constitutes a definitive observation of CISS in a fully nonmagnetic device structure and demonstration of its ability to generate spin accumulation in a conventional semiconductor. The results thus place key constraints on the physical mechanism of CISS and present a new scheme for magnet-free semiconductor spintronics.

**Introduction**

Controlled generation of spin polarization in semiconductors (SCs) is of broad interest for the underlying physics and spintronics and quantum information science applications. A key ingredient for such applications is efficient electrical spin generation in a nonmagnetic SC and transduction of the resulting spin accumulation/current to electrical *signals* (*1,2*). Charge-spin interconversion is commonly realized by contacting the SC with a ferromagnet, which serves as the spin injection source and a spin detector (*3-5*), where the experimental implementation of the spin detection typically takes the forms of a spin-valve or Hanle effect device (*6-8*). In the meantime, nonmagnetic pathways for spin generation and detection have attracted increasing interest. These have included the spin Hall effect (*9,10*) and Edelstein effect (*11*) for charge-to-spin conversion, and their inverse effects (*12*) for spin-to-charge conversion. These effects rely on spin-orbit coupling, and in the case of the Edelstein effect, inversion symmetry breaking (the Rashba effect) (*13,14*) or topological surface states (*15-17*), exploiting helical spin textures in momentum space. In these schemes, the resulting spin polarization is always orthogonal to the direction of the charge current.

More recently, a new nonmagnetic pathway of charge-to-spin conversion has emerged. The effect, termed chirality-induced spin selectivity (CISS), originates from the interplay of electron orbital motion and structural chirality in real space (*18-22*). It manifests as an induced spin polarization in a chiral medium collinear with the charge current along the chiral axis (*19,23,24*). The chirality-induced spin polarization was first evidenced in photo-emitted electrons passing through a self-assembled monolayer (SAM) of double-stranded DNA, via direct Mott polarimetry measurements of the photoelectrons in free space (*25*). In contrast, in solid-state devices incorporating CISS, the spin detection is usually indirect, for instance, by measuring the spin-valve effect using a ferromagnet counter-electrode (*26-36*). Despite the extensive research and preponderance of experimental results, even in the simplest device structures of two-terminal spin-valves, the interpretation of the experiments is fraught with controversy (*21,22*). The open questions include spin versus orbital polarization (*20,37*), the origin of the spin-orbit coupling (SOC) necessary for producing spin polarization (*19,21,35*), and the possible relevance of spinterface effects (*38*) in order to account for the extraordinarily large magnetoresistance (MR) observed in many experiments (*19,21,33,36*). In fact, the very existence of MR in the two-terminal spin valves has been questioned because of its apparent conflict with the Onsager relation (*39-42*). Answers to these questions thus have direct implications for the understanding of the physical mechanism of the CISS effect, which remains elusive (*21,22*). Therefore, for definitive elucidation of both the physical origin and device manifestations of CISS, direct measurements of the polarized spins in robust solid-state/chiral-molecule hybrid devices are imperative (*43*).

Here, we present direct experimental evidence for CISS-induced spin accumulation in a conventional nonmagnetic semiconductor. The spin accumulation is created in a Si-doped GaAs via charge current injection from a Au electrode through an α-helix L-polyalanine (AHPA-L) SAM, and detected via measurement of the Hanle effect without using a magnetic electrode. The Hanle effect data obtained from different devices and the full bias current-temperature parameter space are shown to collapse onto a single scaling function. Notably, the Hanle signals in the *n*-GaAs/AHPA-L/Au junctions follow a distinct power-law temperature ($T$) dependence and a nonmonotonic log-normal-like dependence on the bias current. The bias dependence qualitatively resembles that in conventional ferromagnet/semiconductor devices (*44-47*), while the power-law *T*-dependence with varying onset temperatures may reflect the combination of CISS spin injection via the chiral molecules and spin relaxation in the semiconductor. The experiments thus present an unambiguous case for structural chirality-induced spin generation. The observation of CISS effect in the devices free of any magnetic element places several specific constraints on its theoretical description and suggests a new scheme for magnet-free semiconductor spintronics.

## Results

### Spin accumulation measurement

A schematic diagram depicting the molecular junction device structure and setup for the Hanle measurements is shown in Fig. 1a. The epitaxial layer of Si-doped GaAs was grown by molecular beam epitaxy (MBE) on a semi-insulating GaAs (001) substrate. The carrier (electron) density was determined to be $7.1 \times 10^{18}$ cm$^{-3}$ at 5 K and below from Hall measurements (details in SI 1). The device fabrication process, including the formation and preservation of the AHPA-L SAM, is essentially the same as that employed for making the molecular spin-valve devices based on (Ga,Mn)As (*32,35*). We recently demonstrated that the (Ga,Mn)As/AHPA-L/Au molecular junctions consistently yield pronounced spin-valve magnetoconductance (*32,35*), indicating that the semiconductor-based chiral molecular junctions are a robust and reliable platform for measuring spin-selective transport through chiral molecular SAMs. Most notably, the use of a semiconductor substrate effectively mitigates electrical shorting through defects in the molecular SAM, which is typically fatal in all-metal molecular junctions. Therefore, we expect that replacing the (Ga,Mn)As with a heavily *n*-doped GaAs will produce a molecular junction of similar characteristics, which enables spin detection via the Hanle effect without a magnetic spin analyzer.

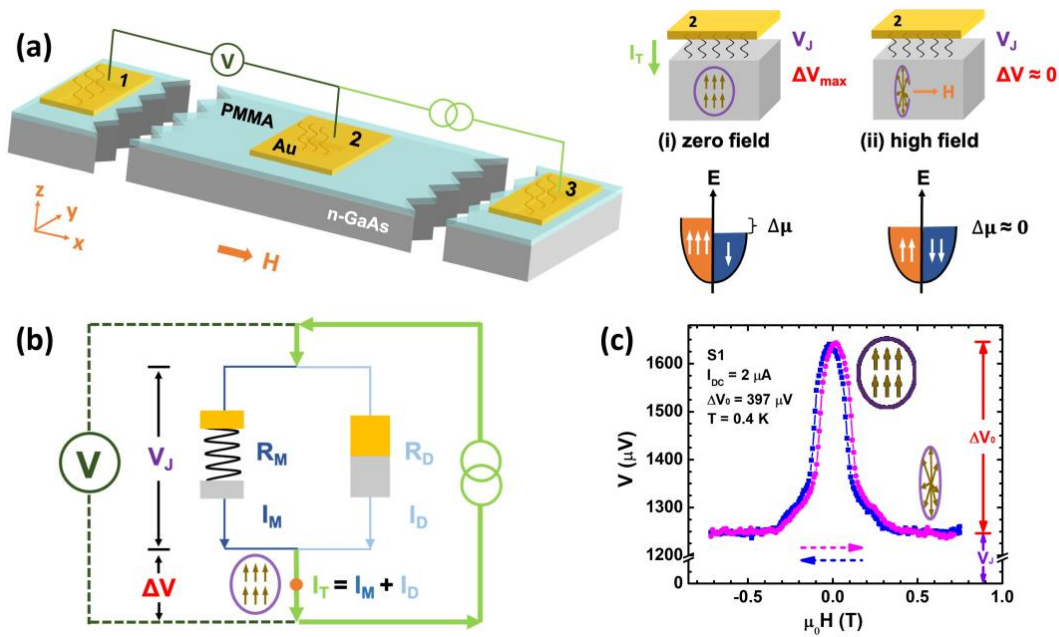

**Fig. 1: Device schematics and representative Hanle curve.** (a) Schematics of the *n*-GaAs/AHPA-L/Au junctions and the measurement setup. A current is applied between contacts 3 to 2, and voltage is measured between contacts 2 and 1. Here, the total voltage measured, $V$, is the sum of junction voltage $(V_J)$ and the spin accumulation voltage $(\Delta V)$. The diagrams on the right depict two different spin states in a chiral molecular junction under electrical current injection: (i) At zero field, the spin polarized current produces spin accumulation in GaAs under contact 2, resulting in spin-splitting of the chemical potential $(\Delta\mu)$ and additional voltage $(\Delta V)$. (ii) When an in-plane magnetic field is applied, the accumulated spins precess around the field and become fully dephased at sufficiently high field, consequently $\Delta V$ decreases to zero while $V_J$ remains essentially unchanged. (b) An equivalent circuit diagram of the molecular junction. $R_M$

($I_M$) and $R_D$ ($I_D$) are the resistances of (currents through) the molecular and direct contact in the molecular junction, respectively. $I_T$ is the total current through the junction. (c) A representative Hanle curve $V(B)$: The measured total voltage $V$ *versus* in-plane magnetic field, as the sum of $V_J$ and $\Delta V$. The blue and magenta dashed arrows indicate the sweep direction of the magnetic field. $\Delta V_0$ is the value of $\Delta V$ at zero field.

The Hanle measurements were performed in a three-terminal (3T) configuration by applying an in-plane magnetic field (Fig. 1a). A fabricated device typically consists of multiple junctions (3 to 4) of size 5x5 µm² with two large reference electrodes. The experimental details for the device fabrication, including the molecular assembly and electrical measurements, are described in the Methods section. Here, the expectation is that a charge current injected into or extracted from the *n*-GaAs will result in spin accumulation of perpendicular polarization due to the CISS effect in the AHPA-L. The spin accumulation, in the form of a spin splitting of the chemical potential, $\Delta\mu = \mu_\uparrow - \mu_\downarrow$, is detected as an additional voltage ($\Delta V$) between the Au and *n*-GaAs electrodes in series with the normal voltage drop across the junction ($V_J$). Upon application of an increasing in-plane magnetic field, the spin accumulation is diminished due to the resulting precession and dephasing of the perpendicularly polarized spins.

Figure 1b shows an equivalent circuit diagram of the chiral molecular junction and a schematic illustration of the expected experimental outcome from the picture of chirality-induced spin injection and accumulation. First, we note that the SAMs in the junctions are most likely not perfect *(48-50)*; defects are almost always present in SAMs at such device scales (µm) (*51*). As a result, electron transport through the junctions comprises two parallel contributions to the total charge current ($I_T$): one through the chiral molecules, $I_M$, and the other through the pinholes in the SAM (direct contact between the Au and *n*-GaAs), $I_D$. The scenario is similar to that in (Ga,Mn)As spin-valve devices (*32,35*). Here, $I_M$ is spin polarized and induces spin accumulation in the GaAs, while $I_D$ does not. The total voltage measured then consists of two components: $V = V_J + \Delta V$, where $V_J$ is the voltage drop across the junction and $\Delta V$ is the spin accumulation voltage in the GaAs (due to $\Delta\mu$). Upon application of an in-plane magnetic field, the magnitude of $\Delta V$ is expected to decrease with increasing field and reach zero, concurrently the measured voltage $V$ becomes constant if $V_J$ has negligible dependence on applied magnetic field.

Figure 1c shows a representative measurement of the total voltage as a function of the applied in-plane magnetic field for a *n*-GaAs/AHPA-L/Au junction. The resulting $V(B)$ curve is qualitatively consistent with the expectations described above: The measured voltage is maximum at zero field and decreases with increasing $B$, reaching a constant value at ~300 mT. The $V(B)$ curve is not exactly Lorentzian and shows small hysteresis in the field sweeps; both features are discussed below.

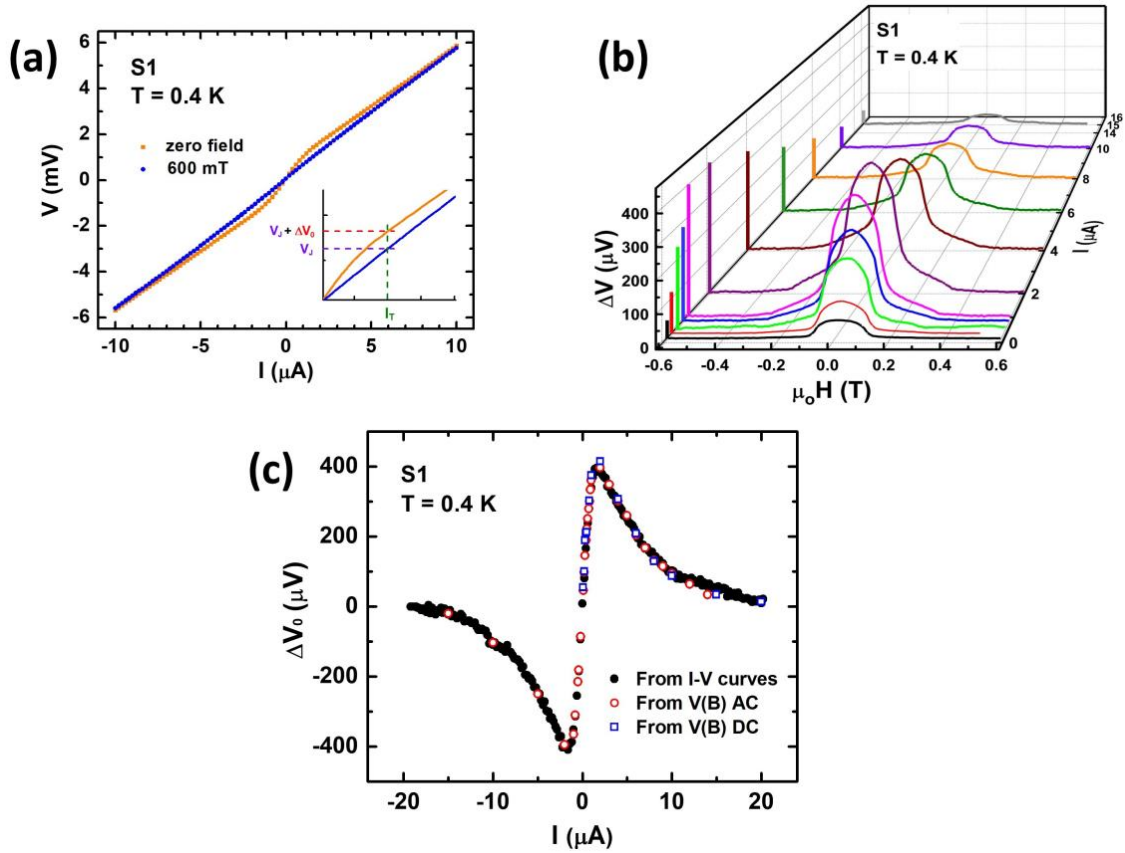

**Fig. 2: I-V characteristics and bias-dependent Hanle signals.** (a) I-V characteristics of the $n$-GaAs/AHPA-L/Au junction in zero (orange) and in-plane magnetic field of 600 mT (blue). The inset shows a close-up of the I-V curves, which depicts how $\Delta V_0$ at a fixed current is extracted from the I-V. (b) Spin accumulation voltage versus applied field measured at different AC bias currents at 0.4 K. (c) The Hanle amplitude, $\Delta V_0$, defined as the spin accumulation voltage at zero field, extracted from I-V curves in (a), AC $V(B)$ curves in (b), and DC $V(B)$ measurements (Fig. S2), as a function of bias current.

Figure 2a shows the I-V curves for the same junction measured at 0.4 K in zero and 600 mT in-plane field, and the result is in full agreement with the $V(B)$ data: At 600 mT, above the saturation field in $V(B)$, the I-V is linear, indicating an Ohmic junction resistance that has negligible dependence on the applied field. At zero field, the I-V shows "nonlinear" behavior in the low-bias regime. However, there is compelling evidence that *the apparent "nonlinearity" is not intrinsic magnetoresistance of the junction, but rather an additional voltage due to spin accumulation* in series with the Ohmic junction: i) A moderate magnetic field of 300 mT eliminates the "nonlinearity" and restores the linear I-V for the junction. ii) With increasing bias current, the zero-field I-V approaches that in the 600-mT field and becomes linear. From the two I-V curves in Fig. 2a, we can extract the amplitudes of the Hanle signals, *i.e.*, the magnitudes of the additional voltage in zero field, $\Delta V_0$, at different bias currents, as illustrated in the inset of Fig. 2a. The same Hanle amplitudes can also be determined from DC and low-frequency AC measurements of $V(B)$ at different fixed bias currents. A set of such measurements on the same junction with different AC currents is shown in Fig. 2b. The resulting Hanle amplitudes are plotted in Fig. 2c; the different measurements produce essentially identical results. $\Delta V_0$ shows a nonmonotonic dependence on the bias current; it initially increases sharply with increasing current and then decreases precipitously and vanishes at higher currents. Another notable feature in Fig. 2a,c is that $\Delta V_0$ is exactly antisymmetric upon reversal of the bias current, which is corroborated by the fact that the DC and AC measurements produce the same results.

We have fabricated and evaluated ca. 20 such devices, each having 3-4 junctions. 6 devices yielded 11 chiral molecular junctions exhibiting signals qualitatively similar to those shown in Fig. 2 out of 39 junctions measured. The majority of the results shown here are from 4 junctions from 4 different samples, on which full sets of temperature- and bias-dependent measurements were performed. Importantly, there were distinct and easily identifiable "failure mode" for the chiral molecular junction devices that did not show Hanle signal (details in SI 3). Similar measurements were also performed for control samples without any chiral molecules. Neither the I-V nor the $V(B)$ showed any discernible magnetic field dependence (details in SI 4). The control experiments provide direct evidence supporting the conjecture that the current through the defects in the chiral SAM ($I_D$) does not contribute to the spin accumulation (Fig. 1b).

**Universal temperature and bias current dependences**

We now examine the general behavior of the Hanle amplitude, $\Delta V_0$, in the temperature-bias current space. Figure 3a shows the $\Delta V(B)$ curves for the junction in Fig. 1 taken at various temperatures with a bias current of 2 µA. Evidently, $\Delta V_0$ decreases with increasing temperature and vanishes at ~6.3 K in S1. Similar measurements were carried out for varying bias currents on both sides of the peak of the $\Delta V_0(I)$ (Fig. 2c); the resulting $\Delta V_0(T)$ for different bias currents are plotted in Fig. 3b. Interestingly, $\Delta V_0$ follows the same $T$-dependence for all bias currents with the same onset temperature $T_0$, as shown by the collapse of all the $\Delta V_0(T)$ data in Fig. 3b onto a single curve by scaling of $\Delta V_0(T)$ for different currents with their respective 0.4 K values, $\Delta V_0(0.4\,K)$, as shown in Fig. 3c. All the scaled data are well described by:

$$\frac{\Delta V_0(T)}{\Delta V_0(0.4K)} = 1 - (T/T_0)^{5/2}. \tag{1}$$

Moreover, the $\Delta V_0(T)$ for different samples can also be scaled to a single function of the reduced temperature, $T/T_0$. Figure 3d shows $\Delta V_0(T)$ for three different samples measured at their respective current of peak $\Delta V_0$ value, $I_p$. The magnitudes of $\Delta V_{0p}$ for the three junctions differ greatly, 398 µV, 144 µV, and 11 µV for S1, S4, and S2 respectively. Curiously, the onset temperatures of the Hanle signals for the junctions exhibit similarly large variations, which appear to correlate with their magnitude, at 6.3 K, 2.3 K, and 1.2 K, respectively. Despite the large variations of $\Delta V_0$ and $T_0$, the normalized $\Delta V_0(T)$ for the three junctions show the same dependence on the reduced temperature $T/T_0$, as shown in Fig. 3e. Here the data for different samples in Fig. 3d are scaled with their respective zero-temperature values and onset temperatures. The $\Delta V_0(0K)$ and $T_0$ values are obtained by fitting each curve in Fig. 3d to Eq. (1). Scaling with $\Delta V_0(0K)$, instead of $\Delta V_0(0.4\,K)$, is necessary for S4 and S2 because of their relatively low $T_0$.

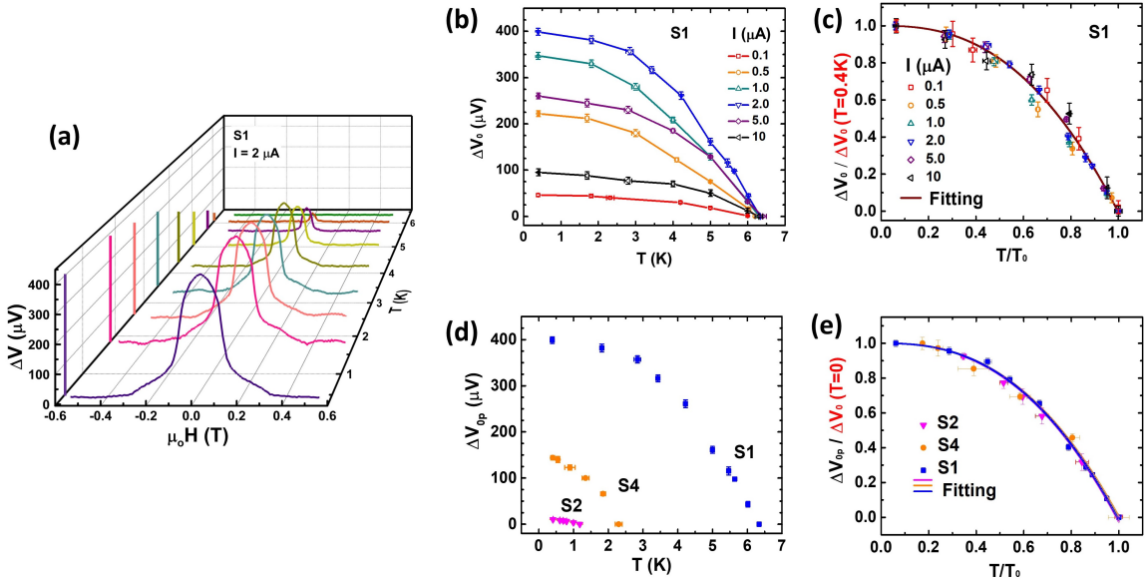

Fig. 3: **Universal temperature dependence of the Hanle amplitude.** (a) $\Delta V(B)$ for S1 at different temperatures measured at a fixed AC bias current of 2 µA. (b) Temperature dependence of the Hanle amplitude, $\Delta V_0$, at different AC bias currents. (c) Scaling of the $\Delta V_0(T)$ data for different bias currents in (b) with its values at 0.4 K and the onset temperature $T_0$. The solid line is a fit to Equation (1). (d) Temperature dependence of the peak Hanle amplitude, $\Delta V_{0p}$, for three different samples. (e) Scaling of the curves in (d) with their respective zero-temperature values and onset temperatures. The $\Delta V_0(0K)$ and $T_0$ values are obtained by fitting each curve in (d) to Equation (1). The solid lines are the fittings to Equation (1) with scaling.

The bias current dependences of the Hanle signals at different temperatures and in different samples show similar universality. Figure 4a shows the variation of $\Delta V_0$ with bias current for junction S1 at different temperatures. With increasing temperature, $\Delta V_0$ decreases over the entire bias range, in such a way that the bias dependence remains unchanged; most notably, the peak current $I_p$ stays essentially constant with changing temperature. Figure 4b shows the data in Fig. 4a normalized by their respective peak values of $\Delta V_0$, $\Delta V_{0p}$. The data at all temperatures collapse onto a single curve, indicating a common bias-dependence that is independent of temperature. Moreover, for different samples, while the peak values of $\Delta V_0$ and the peak currents vary greatly (Fig. 4c), all the $\Delta V_0(I)$ data once again fall onto a single curve when $\Delta V_0$ and bias current are normalized by their respective peak values; the scaling behavior is evident in Fig. 4d.

The consistency of the experimental results across the different junctions and a common underlying temperature-bias current dependence have a striking manifestation, shown in supplementary Figure S5: The scaled results show that the signal at 0.4 K in junction S2 closely resembles the signal at 5.4 K for junction S1 (details in SI 5).

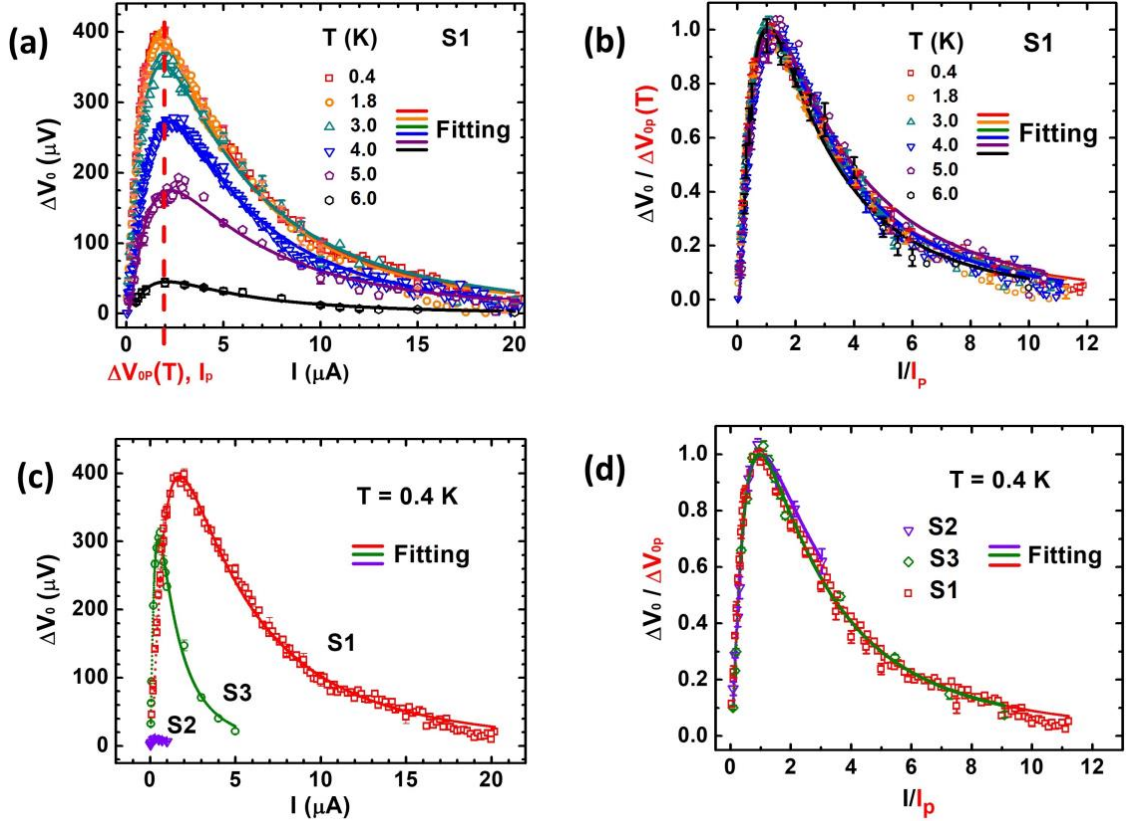

Fig. 4: **Universal bias current dependence of the Hanle amplitude.** (a) Bias current dependence of $\Delta V_0$ at different temperatures for S1. The magnitude of $\Delta V_0$ first increases then decreases with the bias current. The peak value of $\Delta V_0$ is defined as $\Delta V_{0p}$ at the current of $I_p$. (b) Scaling of the $\Delta V_0(I)$ data in (a) for different temperatures with their respective values of $\Delta V_{0p}$ and $I_p$. (c) Bias current dependence of $\Delta V_0$ for three different samples at 0.4 K. (d) Scaling of the data in (c) with their respective values of $\Delta V_{0p}$ and $I_p$. The solid lines in all panels are best fits to Equation (4).

## Discussion

### Three-terminal Hanle measurement

The Hanle effect has been one of the most widely utilized and effective methods of measuring spin lifetime, spin accumulation, and spin transport in SCs. The measurements have been implemented in both nonlocal four-terminal (NL-4T) (*7,8*) and simplified three-terminal (3T) geometries (*4,52*). A notable controversy in the field concerns the reliability of 3T Hanle measurements for spin detection (*52*). Specifically, in devices with oxide barriers, it was shown that i) the amplitudes of the 3T Hanle signals often greatly exceeded the values expected from the Valet-Fert theory (*53,54*) and NL-4T measurements (*55,56*) and, ii) the spin lifetimes determined from the 3T Hanle curves were consistently 1-3 orders of magnitude smaller than theoretical predictions (*57*) and results of electron spin resonance (*58*) and NL-4T measurements (*52*). This has led to critical questions and conclusions that the 3T Hanle measurement does not probe spin injection and accumulation. However, it is important to note that there is no fundamental physical reason prohibiting spin injection and detection in the 3T Hanle setup. Moreover, a ubiquitous component of the 3T Hanle devices showing the anomalous properties is an artificial oxide tunnel barrier, which hosts the localized electronic states necessary in the alternative models proposed to account for the anomalous observations (*53,59*). In contrast, in all-epitaxial heterostructure devices free of oxide barriers, the 3T Hanle results do not show the anomalous behaviors described above (*45,60,61*), but rather are fully consistent with the results of NL-4T Hanle measurements in the same devices. We recently directly compared 3T and NL-4T Hanle measurements in the same devices of epitaxial Fe/Al$_{0.3}$Ga$_{0.7}$As heterostructures; the 3T Hanle signals exhibit broad similarities with the NL-4T results in all aspects, including similar spin lifetimes and consistent Hanle amplitudes (*45*). The experiments provided compelling evidence that in devices engineered to minimize localized states in spin injectors and detectors, the 3T Hanle effect reliably probes spin accumulation and its dynamics in the semiconductor channel.

For the molecular junction devices studied in this work, a direct comparative NL-4T Hanle measurement is exceedingly difficult to implement because a lithographical step would be required after the molecular assembly, which would destroy the AHPA-L SAM. Nevertheless, many aspects of the 3T Hanle data, including the field-suppression of the I-V nonlinearity discussed above and the resulting physical parameters (e.g., spin lifetime) presented below, point to spin accumulation in the *n*-GaAs as the physical origin of the 3T Hanle effect.

**Spin accumulation measurement**

The results in Figs. 3c,e and 4b,d indicate that the amplitude of the Hanle signal varies independently with temperature and bias current. Analytically, the Hanle amplitude $\Delta V_0$ at temperature $T$ and bias current $I$ for any sample can be described by the same expression:

$$\Delta V_0(T, I) = \Delta V_{max} f(T) g(I), \qquad (2)$$

where,

$$f(T) = 1 - \beta (T/T_0)^\alpha, \qquad (3)$$

and

$$g(I) = exp\left(-\frac{[ln(I/I_p)]^2}{w}\right). \qquad (4)$$

Several features are worth noting: i) $T_0$ is independent of $I$; ii) $I_p$ is independent of $T$; iii) $\Delta V_{max} = \Delta V_0(0, I_p)$, $T_0$, and $I_p$ are sample-specific and determined directly from experiment; iv) both $T_0$ and $I_p$ show positive correlation with $\Delta V_{max}$. The best fits of the data in Fig. 3c,e to Eq. 3 yield $\beta$ of unity (1±0.02) and $\alpha$ of 5/2 (details in SI 6). The temperature dependence in Eq. 3 resembles that of the magnetization of an isotropic ferromagnet described by the Bloch's law (*62*), but with a power law exponent of 5/2 instead of 3/2. A rendition of $\Delta V_0(T, I)$ in the full temperature-bias current space for S1 is shown as a false-color plot derived from Eq. (2) in Fig. S8.

The striking nonmonotonic bias current dependence of the Hanle signals can be quantitatively described by the log-normal-like function of Eq. 4. The unscaled data in Fig. 4a,c are fit to Eq. 2, and the best fits are shown as the solid lines. Here $w$ is practically the only adjustable fitting parameter, as $\Delta V_{max}$ and $I_p$ are obtained directly from the experimental data with high accuracy. Interestingly, the resulting $w$ from the three different samples in Fig. 4c are essentially the same (2.26 ± 0.04), despite the large differences in the Hanle amplitude and peak current. The $w$ derived from junction S1 at different temperatures (Fig. 4a) shows a slight, but noticeable, decrease with increasing temperature near $T_0$. More details of the fitting procedures are given in SI 6. The values of the physical and fitting parameters for the same sample at varying temperature and for different samples at the same temperature are listed in Tables S2 and S3, respectively (details in SI 8).

We now turn to the physical implications of the results outlined above. First, we summarize the key observations from the Hanle measurements: i) The Hanle amplitude follows a nonmonotonic log-normal-like dependence on the bias current for all samples. ii) The Hanle amplitude exhibits a well-defined power-law decrease with increasing temperature at all bias currents for all samples. iii) The onset temperature ($T_0$) of the Hanle signal is sample dependent, and correlates positively with its maximum amplitude.

For the bias dependence, we first note the spurious nature of the peak current value, $I_p$, due to the presence of parallel conduction current through the direct contact. Therefore, any correlation of $I_p$ with $\Delta V_{max}$ must be viewed with caution. Nevertheless, we emphasize that the parallel conduction current has no bearing on the bias-dependence, because of the intrinsic linear I-V characteristics of the junctions; both $I$ and $I_p$ through the molecular SAM differ from their respective total values through the junction by the same constant scaling factor, thus not affecting $I/I_p$.

Note that the close similarities of the bias dependence of the Hanle signals here and that of the spin accumulation in conventional ferromagnet/semiconductor (FM/SC) devices (*44-47,63*). Most relevant are the cases of 2T and 3T FM/SC devices, in which the bias dependences of local spin accumulation have been measured both electrically (*45,47*) and optically (*44,63*). Electrically, the spin accumulation was measured via spin-valves (*47*) and the Hanle effect (*45*); in both cases, the magnitudes of the spin accumulation were observed to exhibit similar nonmonotonic bias dependences. The nonmonotonic behavior of the spin accumulation was attributed to a combination of linear increases with injection current and exponential decreases of the spin polarization, which was demonstrated explicitly by Fujita *et al.* (*47*). Optically, the spin accumulation was determined from the circular polarization of the electroluminescence in FM/SC spin LEDs (*63*). Notably, from the electroluminescence of the current injection from Fe into *n*-GaAs, Hickey *et al.* (*44*) identified a bias-dependent polarization at low temperature that resembles our observations in the chiral molecular junctions (Fig. 4). In all cases, the steep drop-off at high biases was attributed to enhanced Dyakonov-Perel (DP) spin relaxation with increasing kinetic energy (momentum) of the injected electrons, for the DP process is the dominant spin relaxation mechanism at high doping densities (*64*).

Evident from the universality of the bias dependence in Fig. 4 is that the value of *w* is essentially the same for all samples despite the large differences of the peak Hanle amplitude. The constancy of *w* is strongly suggestive that the bias dependence reflects the spin dynamics in the *n*-GaAs and is likely independent of the details of the molecular junctions. In this respect, our previous study of the Hanle effect in epitaxial Fe/*n*-AlGaAs devices provided a useful reference (*45,46*). The detailed comparison of molecular junction *n*-GaAs/AHPA-L/Au with epitaxial Fe/*n*-AlGaAs is presented in SI 9. As shown in Fig. S8c, the value of *w* decreases continuously with the carrier density of the *n*-AlGaAs and smoothly approaches the value 2.3±0.1 in the *n*-GaAs in this work. This consistent carrier density dependence of *w* further attests to the common underlying physical processes in our magnet-free chiral molecular junctions and conventional FM/SC devices.

In contrast, the distinct temperature dependence of the Hanle signals observed in our chiral molecular junction devices cannot be understood based solely on spin relaxation in *n*-GaAs. The magnitude of spin accumulation in a normal conductor under spin injection is approximated well by the simplified Valet-Fert equation: $\Delta V/j = \gamma^2 r_N$ (*65*), where $j$ is the injection current density, $\gamma$ is the spin injection/detection efficiency, and $r_N$ is the spin-resistance of the normal conductor (details in SI 10). In our case, the normal conductor is degenerately doped *n*-GaAs, in which the temperature dependence of $r_N$ originates from that of the spin lifetime, $r_N \sim \sqrt{\tau_s}$ (details in SI 11). Theoretically, different power-law *T*-dependences were predicted based on different spin relaxation mechanisms (*2,66*), including $T^{-5}$ from the Elliot-Yafet process (*67*). However, such spin relaxation depends primarily on spin-lattice interactions, which are expected to saturate at low temperatures; more importantly, $\sqrt{\tau_s}$ would yield a direct power-law decrease of spin accumulation with increasing temperature, instead of that of Eq. (3). Experimentally, in conventional FM/SC structures, the spin accumulation signals in *n*-GaAs were observed to exhibit rather weak temperature dependences, persisting above 100 K (*6*). This result is in contrast to the steep power-law decrease and low onset temperatures observed in the chiral molecular junctions in this work. Most notably, as is evident in Fig. 3c, the Hanle signal vanishes at rather low temperatures, and the onset temperatures of different junctions show distinct correlation with their peak amplitudes. These features suggest that the observed *T*-dependence predominantly originates from that of γ. We note that in conventional magnetic tunnel junctions, a similar power-law decrease (with an exponent of 3/2) of the magnetoconductance ($\Delta G$) was observed (*68*). The *T*-dependence there was ascribed to the variation of spin polarization due to thermally excited spin waves. More interestingly, despite the common *T*-dependence in different devices, it was observed that both the amplitude of spin polarization (the equivalent of $\Delta V_{max}$ in this work) and the critical temperature (the equivalent of $T_0$) were sensitive to the interface quality (*68*), which are similar to the behaviors of our junctions, shown in Fig. 3d. The well-defined *T*-dependence in our chiral molecular junctions thus likely reflects that of the mechanism of spin selectivity, whose understanding should provide important new insight into the physical origin of CISS.

Another intriguing feature of our results is the remarkably small bias currents for producing the sizable spin accumulation and for inducing spin relaxation ($I > I_p$). For comparison, these current densities are ca. two orders of magnitude smaller than those in similar conventional SC/FM devices (*6,45*). This result is in spite of the likely presence of parallel conduction in the junctions. The anomalously small current densities are also reflected in the unphysical values of γ inferred from the Valet-Fert model (*65*). For instance, based on the low-current ($I < I_p$) data in Fig. 2c, the Valet-Fert model yields $\gamma \sim 10$ (for details, see SI 10). We surmise that this result reflects the discrete nature of the electron transport through individual chiral molecules; namely, the effective current density is much greater than the average value obtained using the entire junction area, supporting the proposition that CISS is a single-molecule effect rather than a collective effect (*69,70*).

Finally, we examine the line-shape of the Hanle signals. The electrical Hanle effect, which can involve spin precession, spin relaxation, and spin drift/diffusion, provides rigorous elucidation of spin dynamics via analysis using the spin drift-diffusion equation (*61,64,66*). In the absence of spin drift or diffusion or both, the Hanle curve is approximated well by a Lorentzian function (*4,7*), whose full width at half maximum (FWHM) provides a convenient measure to determine the spin lifetime. In our devices, the spin lifetimes estimated from the widths of the Hanle curves range from 100 to 300 ps (details in SI 12), which is in general agreement with the values expected in GaAs at such carrier densities (*64*). The line-shapes of the Hanle signals in the chiral molecular junctions are not precisely Lorentzian. Various factors may contribute to the deviation of the Hanle signal from a pure Lorentzian (*4,66*). First, we note that the Hanle amplitudes are not affected by the line-shapes; thus, the deviations from the Lorentzian have no impact on the analyses of the bias and temperature dependences of the Hanle effect. Moreover, in our devices, most of the Hanle curves appear to be superpositions of two Lorentzian-like curves. In some cases, the central peak is sharp (*e.g.*, Fig. S11 in SI 13), resembling that observed in conventional Fe/*n*-GaAs devices due to the dynamic nuclear polarization effect (*61*). Confirmation of the dynamic polarization of the long-lived nuclear spins in the semiconductor via CISS in the chiral molecular junctions would open a new pathway for fundamental studies of CISS and provide a platform for integral quantum information storage (*22,71*).

Utilizing the robust device platform of semiconductor-based chiral molecular junctions, we demonstrated CISS-induced spin accumulation in a conventional semiconductor and its direct detection via the Hanle effect. The Hanle effect follows distinct universal temperature and bias current dependences. We anticipate using well-defined, quantitative experimental results to elucidate the physical mechanism of CISS. Practically, the successful incorporation of CISS into a fully nonmagnetic device architecture presents a new scheme of magnet-free semiconductor spintronics.

**Acknowledgements**

We acknowledge helpful discussions with Profs. Vladimiro Mujica and Julio L Palma Anda. The work at FSU is supported by NSF grants DMR-1905843 and DMR-2325147. The work at UCLA is supported by NSF grant CHE-2004238 and the W. M. Keck Foundation through the Keck Center on Quantum Biology. The work at IOS is supported by the MOST grant 2021YFA1202200, the CAS Project for Young Scientists in Basic Research (YSBR-030) and the Strategic Priority Research Program of the Chinese Academy of Sciences under Grant No. XDB44000000.

0. Methods

1) Materials

The AHPA-L in the experiments was obtained from GenScript, LLC. The resistivity of the commercial semi-insulating GaAs (001) is larger than $1 \times 10^7 \ \Omega \cdot cm$. Molecules were dissolved in pure ethanol at 1 mM concentration. The AHPA-L solution was kept at -18 °C for storage.

The Si-doped GaAs substrates were grown by molecular beam epitaxy. A 100 nm-thick GaAs buffer layer was first grown on semi-insulating GaAs (001) at 560 °C. A 400 nm-thick GaAs layer doped with Si was later grown at the same temperature. The results shown in this paper are from two GaAs substrates with carrier concentrations of $5 \times 10^{18} \ cm^{-3}$ and $7.2 \times 10^{18} \ cm^{-3}$.

2) Device fabrication process

The Hanle devices were fabricated in three steps: define junctions by electron-beam lithography (EBL); remove the oxide layer on GaAs and assemble AHPA-L on the junctions; and deposit the top Au electrodes. The parameters are similar to the fabrication procedure of CISS spin-valve devices previously reported (*1,2*).

The junction size is $5 \times 5 \ \mu m^2$ in all devices. The top electrode consists of 5 nm of Cr and 35 to 40 nm of Au. During the evaporation, the sample was mounted on an angled stage of 15° to ensure the coverage and continuity of the metal film at the edge of the junctions. The substrate temperature was maintained between -30 °C and -50 °C via liquid nitrogen cooling. For both the spin-valve devices studied previously (*2*) and the Hanle devices studied in this work, junctions were also fabricated without the Cr adhesion layer and measured. The two types of devices yielded essentially the same results, whereas the junctions with Cr adhesion layers tended to be mechanically more robust.

To obtain the carrier density of the GaAs substrate, Hall electrodes were patterned by photolithography and followed by wet chemical etching on bare GaAs. The GaAs etchant is $H_2SO_4:H_2O_2:H_2O$ (1:8:40) and it etches at a rate of 12.6 nm/s. Most devices were etched for a depth of 430-450 nm.

3) Electrical measurements

In Hanle measurements, samples were fixed on a socket with a copper base and wired by hand with silver paint and Pt wire. In the Hall measurements, contacts were made via indium soldering to reduce the contact resistance. All samples were measured in an Oxford $^3$He cryostat. DC measurements were done with Keithley 2400 as the current source and HP 3458 as the voltmeter. AC measurements were performed with SR2124 dual-phase analog lock-in amplifiers.

In all Hanle measurements presented in this manuscript, the magnetic field was applied parallel to the sample surface.

## 1. Characterization of GaAs Epilayers

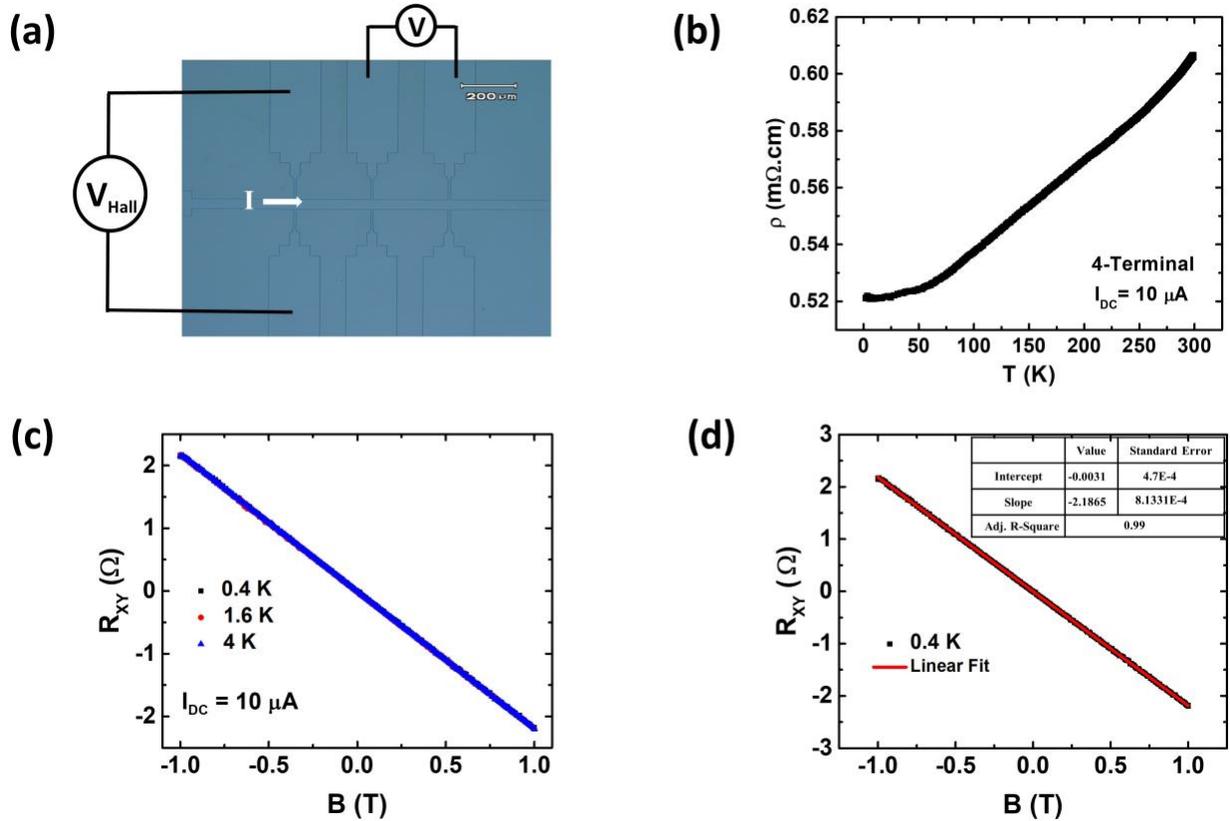

Fig. S1 (a) An optical image of a Hall bar pattern of an epitaxial film of a Si-doped GaAs film on a semi-insulating GaAs substrate. (b) Temperature-dependent resistivity from 4-probe measurements with a DC bias current of 10 µA. (c) Hall resistance of the GaAs film at 0.4 K, 1.6 K, and 4 K. (d) Linear fitting of Hall resistance as a function of perpendicular magnetic field at 0.4 K.

The Si-doped GaAs epitaxial films were characterized using 4-probe resistivity and Hall effect measurements, as depicted in the optical image in Figure S1(a). The sample resistance was measured with a 4T setup at a DC bias current of 10 µA. Figure S1(b) illustrates the temperature dependence of the thin film, exhibiting metallic behavior. This metallicity is consistent with the high doping level (>10$^{18}$ cm$^{-3}$) of Si in the GaAs.

The Hall measurements were performed at 0.4 K, 1.6 K, and 4 K, which show essentially identical results, as shown in Figure S1(c). These results indicate temperature-independent carrier density within this range, as expected for a degenerately doped semiconductor. The linear fit of the Hall resistance as a function of perpendicular magnetic field (Fig. S1(d)) gives the slope $R_B = \frac{V_H}{I_{DC} \times B} = \frac{R_H}{B} = \frac{1}{net} = 2.18 \, \Omega/T$, where $V_H$ is Hall voltage, $I_{DC}$ is the applied DC current, $B$ is the perpendicular applied field, $n$ is the carrier density, $t$ is the thickness of the doped GaAs thin film, and $e$ is the charge of an electron. The resulting carrier density $n$ is $7.1 \times 10^{18} \, cm^{-3}$, consistent with the intended Si doping level in the growth. The carrier mobility was $\mu = \frac{1}{\rho n e} = 1588 \, \frac{cm^2}{V.s}$.

In the experiments, S1, S2, and S3 were fabricated using a GaAs epilayer with a carrier density of $7.1 \times 10^{18} \, cm^{-3}$, while S4 utilizes a GaAs epilayer with a carrier density of $5.0 \times 10^{18} \, cm^{-3}$.

## 2. DC Measurements of the Hanle Effect

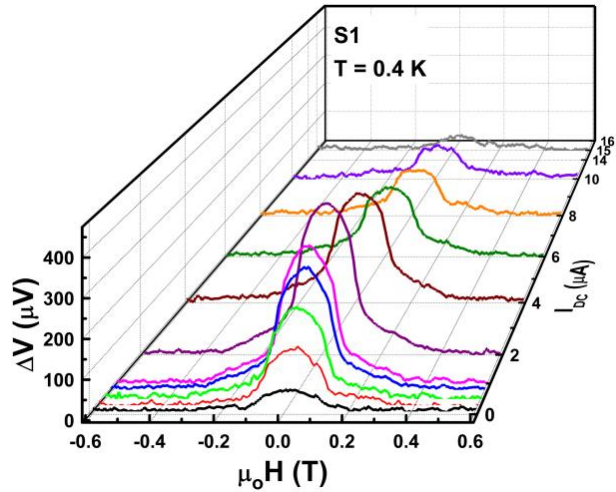

Fig. S2: Spin accumulation voltage *versus* applied field measured as a function of DC bias currents at 0.4 K.

Figure S2 shows the spin accumulation voltage *versus* applied field measured as a function of DC bias current at 0.4 K. The results from DC measurements are similar to the AC measurements, shown in Fig. 2(b). We have included the results from AC measurements in the main manuscript due to their lower noise levels (~5 μV) compared to DC (~15 μV) measurements.

## 3. Distinguishing Characteristics between 'Working' and 'Failed' Junctions

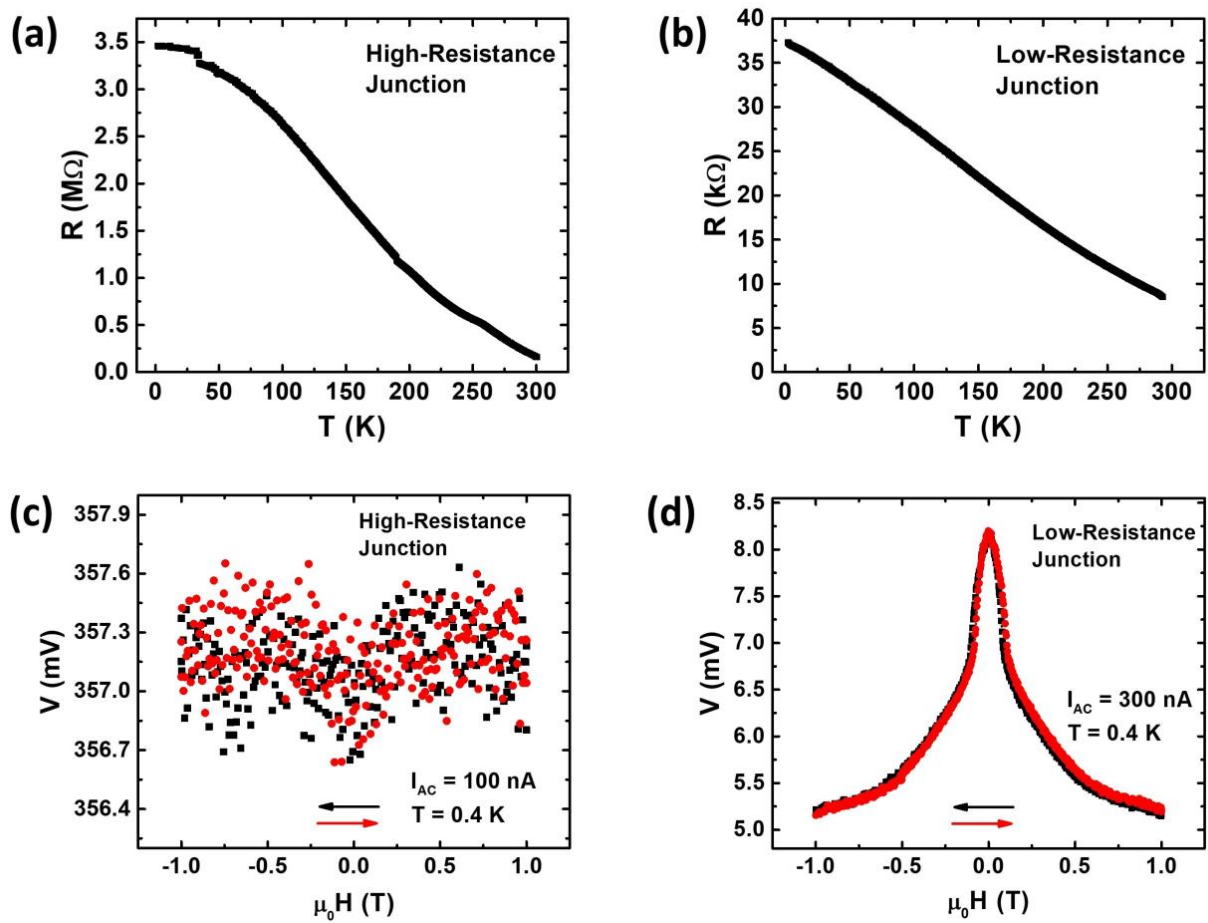

Fig. S3: Representative junction resistance as functions of temperature, $R(T)$, for (a) a 'failed' junction without measurable Hanle signal, and (b) a 'working' junction with Hanle signal. Corresponding Hanle measurement for (c) 'failed' junction and (d) 'working' junction at 0.4 K.

As stated in the manuscript, *ca.* 20 samples, each containing 3-4 junctions, were fabricated and measured in this work. Detailed information regarding the statistics of working/non-working junctions from different samples is tabulated in Table S1. From the large number of devices, we have observed two distinct and easily identifiable features that consistently distinguish the working devices (with measurable Hanle signals) from the 'failed' devices that did not show any measurable Hanle signal: i) The absence of measurable Hanle signal was always associated with a high junction resistance. Figure S3(a) is a typical $R(T)$ plot for a 'failed' junction, where the junction resistance went up to MΩ range at low temperature. We observed that the junction resistances for 'failed' junctions were always larger than hundreds of kΩ. In contrast, as exemplified in Fig. S3(b), the junction resistances for 'good' junctions were a few kΩ to tens of kΩ. ii) The absence of Hanle signal was also associated with a large phase angle (up to 70-80 degrees, in contrast to essentially 0 for 'good' devices in ac lock-in measurements. Both features were evident at the room temperature. The corresponding Hanle measurement of the 'failed' and 'good' junctions at 0.4 K are shown in Fig. S3(c) and (d), respectively.

Most likely, the high junction resistance and large phase angle resulted from poor molecular assembly; in contrast to an ordered SAM, a disordered thicker molecular layer leads to high junction resistance and a large capacitive contribution manifested in a large phase angle. Therefore, 'failed' devices are easy to identify. In our experiments, the primary factor affecting the molecular assembly and resulting device quality was probably variation in the preparation of the GaAs surface by Ar ion milling due to drift in the ion current. Another, more minor, factor is variation of the substrate temperature during deposition of the top metal layer.

| *n*-GaAs Wafer | Devices made from the wafer | Devices with working junctions | Total number of junctions measured | Number of working junctions | Comment |
|---|---|---|---|---|---|
| **Fabricated and Measured: January - March 2020** | | | | | |
| 3926 | 4+1 (Control) | 2 | 12 | 4 | Non-working junctions had large phase angles |
| 3927 | 2 | 0 | 6* | 0 | Insulating substrate |
| 3931 | 1 | 0 | 3* | 0 | Insulating Substrate |
| **Fabricated and Measured: April - August 2022** | | | | | |
| 3930 | 3+1 (Control) | 1 | 8 | 2 | Non-working junctions had large phase angles |
| 3944 | 7+1 (Control) | 3 | 19 | 5 | Non-working junctions had large phase angles |
| **Total** | **17+3** | **6** | **39** | **11** | |

*These junctions were made on insulating *n*-GaAs substrates and are not included in the total number of junctions measured

Table S1: Tabulation of statistics of 'working' and 'failed' junctions.

## 4. Control Experiments without Molecules

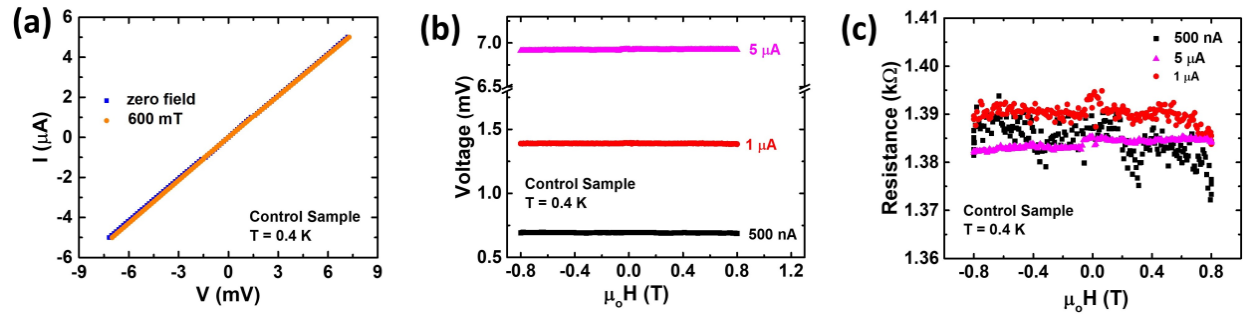

Fig. S4: Control sample measurements: (a) I-V curves for a control junction (Au/GaAs) in zero magnetic field and an in-plane field of 600 mT at 0.4 K. (b) Junction voltage, and (c) junction resistance *versus* in-plane magnetic field with different bias currents.

The control samples were fabricated using the same device fabrication process as the molecular junctions, with the exception that we omitted the assembly of molecules. In Figure S4(a), the I-V curves are presented both without an external magnetic field and with an in-plane field of 600 mT at 0.4 K. As anticipated, the I-V curves are linear and indistinguishable with or without the application of a magnetic field. Figures S4(b) and (c) show that the curves of V or R *versus* B do not exhibit any noticeable magnetic field dependences above the noise level. This observation provides direct evidence supporting the hypothesis that the current passing through defects in the chiral SAM ($I_D$) in the molecular junctions does not contribute to spin accumulation.

## 5. Consistency of the Hanle Signals between Different Samples

The working samples exhibited similar behaviors with regard to bias current and temperature dependences of $\Delta V_0$. Here, we present the experimental results from different samples (excluding sample S1, which is already shown in main manuscript):

A. Bias Current Dependence

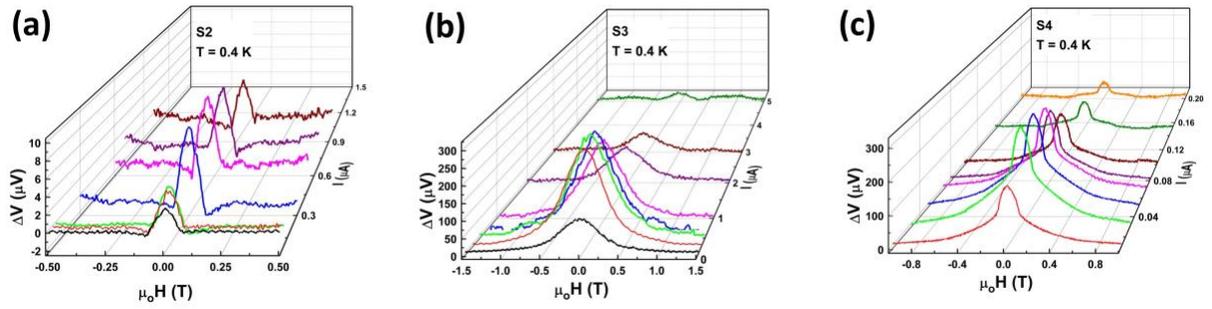

Fig. S5: Hanle curves, ΔV(B), for (a) S2, (b) S3, and (c) S4, measured at different AC bias currents at a fixed temperature of 0.4 K.

Figure S5 shows the Hanle curves, $\Delta V(B)$, for three samples measured at different AC bias currents at a fixed temperature of 0.4 K. While the magnitude of the $\Delta V_0$ varies among the samples, the nature of signals with varying bias current is essentially the same as that observed in sample S1 (Fig. 2(b)). The bias current dependence of all samples exhibits consistent non-monotonic behavior, as depicted in Fig. 4(c) of the main text.

B. Temperature Dependence

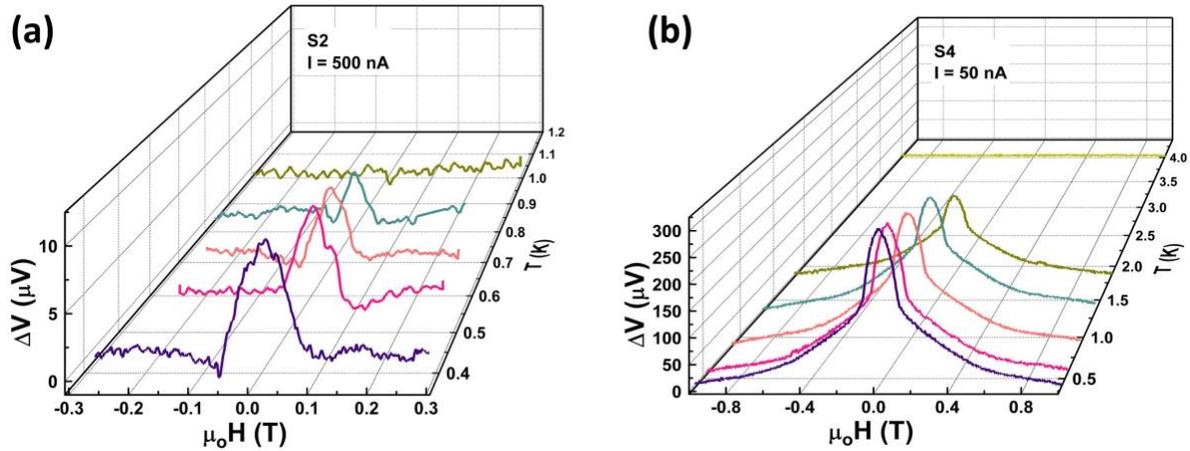

Fig. S6: Temperature dependence of $\Delta V$ measured in (a) S2, and (b) S4 at AC bias current of 500 nA and 50 nA, respectively.

Figure S6 shows the temperature dependence of $\Delta V(B)$ for two samples measured at a fixed current. Note that the onset temperatures differ between the samples. However, the temperature dependences in S2 and S4 are similar to that observed in S1 (Fig. 3 of the main text).

C. Similar Hanle Curves of Different Samples at Different Temperatures

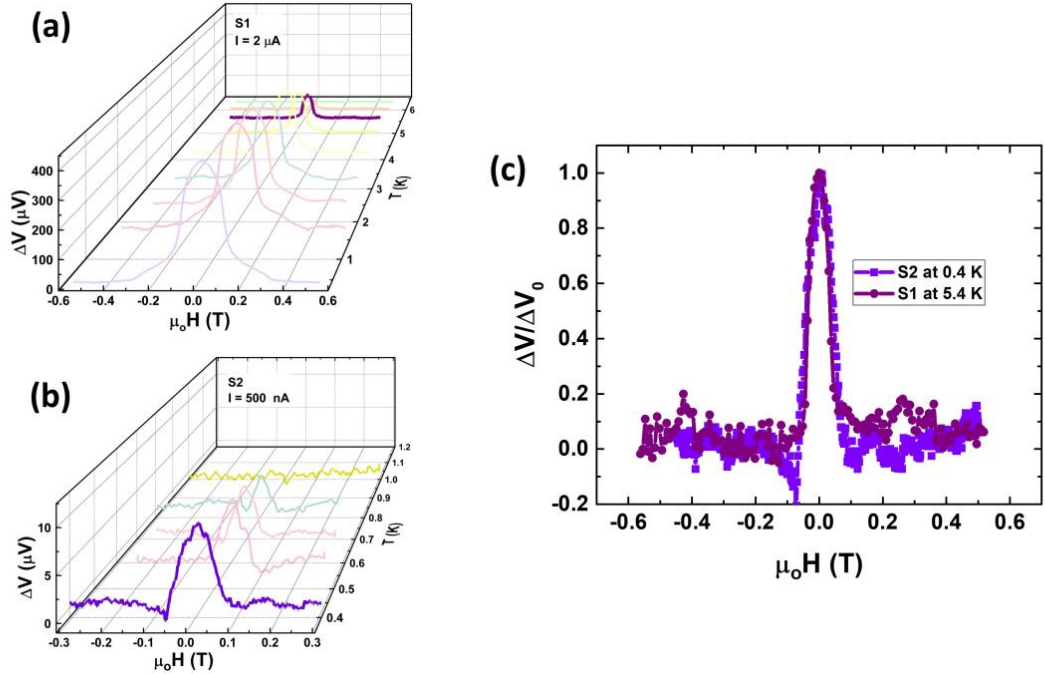

Fig. S7: Similarity between the two Hanle curves (highlighted) from two different samples at different temperatures: (a) S1 at 5.4 K with (b) S2 at 0.4 K. After scaling the amplitude, the two curves are essentially the same, as shown in (c).

The consistency of the experimental results among different devices can be further ascertained by comparing the Hanle curves from different samples. Figures S7(a and b) show the Hanle curves from S1 measured at 5.4 K at a bias current of 2 µA and S2 measured at 0.4 K at a bias current of 0.5 µA, respectively. After scaling the amplitude, the two curves are essentially the same, shown in Figure S7(c).

## 6. Details on Data Rescaling and Fitting Procedures

The scaled results for the temperature and bias current dependences of $\Delta V_0$ were fit using Origin data analysis software.

The temperature dependence of $\Delta V_0$ was first analyzed with a power-law function with two adjustable parameters, α and β:

$$\frac{\Delta V_0}{\Delta V_0(T=0)} = 1 - \beta \left(\frac{T}{T_o}\right)^{\frac{\alpha}{2}} \tag{S1}$$

Figure S8 shows the fitting procedure as follows: First, the data were fit with two variables, where the best fit gives the values of α and β as 4.84 and 0.99, respectively. Subsequently, a secondary fit was performed, fixing the value of α while allowing β to vary (with values as tabulated in table of Fig. S8(a)). The best fit of the data thus yields β of unity (1±0.02) and $\alpha$ of 5. The same procedure was followed for all samples.

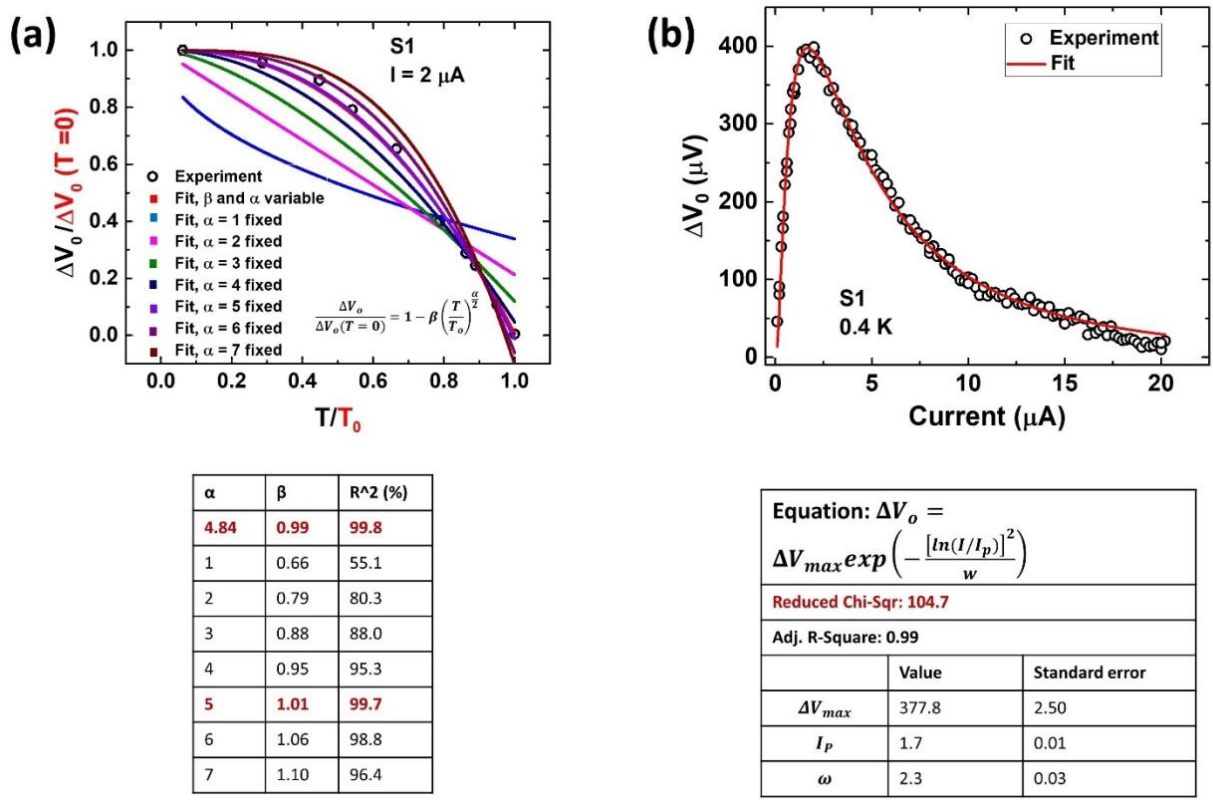

Fig. S8: An example depicting the process of data fitting for the (a) temperature dependence, (b) bias current dependence of $\Delta V_0$ in Origin. Similar processes of fitting were deployed for all samples.

The bias current dependence of $\Delta V_0$ was analyzed with the scaling function:

$$\Delta V_0 = \Delta V_{max} exp\left(-\frac{[ln(I/I_p)]^2}{w}\right) \quad (S2)$$

In this analysis, the data were initially fit to Equation S2, which includes three variables: $\Delta V_{max}$, $I_p$, and $w$. As shown in Fig. S8(b), the best-fit results yielded values for $\Delta V_{max}$ and $I_p$ that align closely with the experimental observations. Note that $\Delta V_{max}$ and $I_p$ are parameters specific to each individual measurement and can be determined directly from the experimental results to high precision. On the other hand, $w$ represents an adjustable fitting parameter in the given equation. Therefore, any variations in the value of $w$ resulting from changes in physical parameters, such as carrier concentration, hold significance in the context of spin accumulation and relaxation processes within the semiconductor GaAs.

## 7. Full Bias and Temperature Dependences

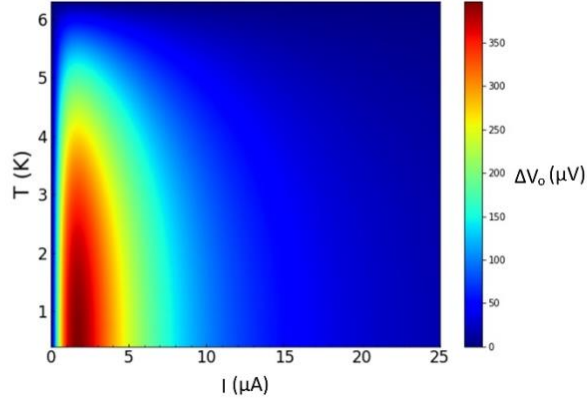

Fig. S9: A false-color plot showing the full bias and temperature dependences of $\Delta V_0$ for S1. Here the data set is generated using Eq. 2 in the main text.

Figure S9 displays a false-color plot for S1, illustrating the full bias and temperature dependences of $\Delta V_0$ within the observable ranges. To create this plot, the experimental values of $\Delta V_{max}$ at 0.4 K and $T_0$ were included, while the remaining data set was generated by applying Equation 2 in the main text.

## 8. Tabulating the Fitting Parameters from Different Samples

Using similar fitting procedures for bias dependence of $\Delta V_0$ as explained in Section 5, any change in the adjustable parameter $w$ was analyzed for different temperatures and samples. The values of various parameters for the same sample at varying temperatures and for different samples at the same temperature are listed in Tables S2 and S3, respectively. The physical significance of these parameters is explained in detail in the main text.

| Temperature (K) | $\Delta V_{max}$ (µV) | $I_p$ (µA) | $w$ |
|---|---|---|---|
| 0.4 | 397.72 | 1.71 | 2.27 |
| 1.8 | 388.76 | 1.75 | 2.25 |
| 3.0 | 353.38 | 1.78 | 2.23 |
| 4.2 | 274.47 | 1.81 | 2.12 |
| 5.0 | 174.98 | 1.87 | 2.05 |

Table S2: Fitting parameters for sample 1 at different temperatures.

| Sample | $V_{max}$ (µV) | $I_p$ (µA) | $w$ |
|---|---|---|---|
| 1 | 397.72 | 1.71 | 2.27 |
| 2 | 10.57 | 0.32 | 2.23 |
| 3 | 302.98 | 0.50 | 2.29 |

Table S3: Fitting parameters for different samples.

## 9. Comparisons between Au/AHPA-L/n-GaAs and Epitaxial Fe/n-AlGaAs

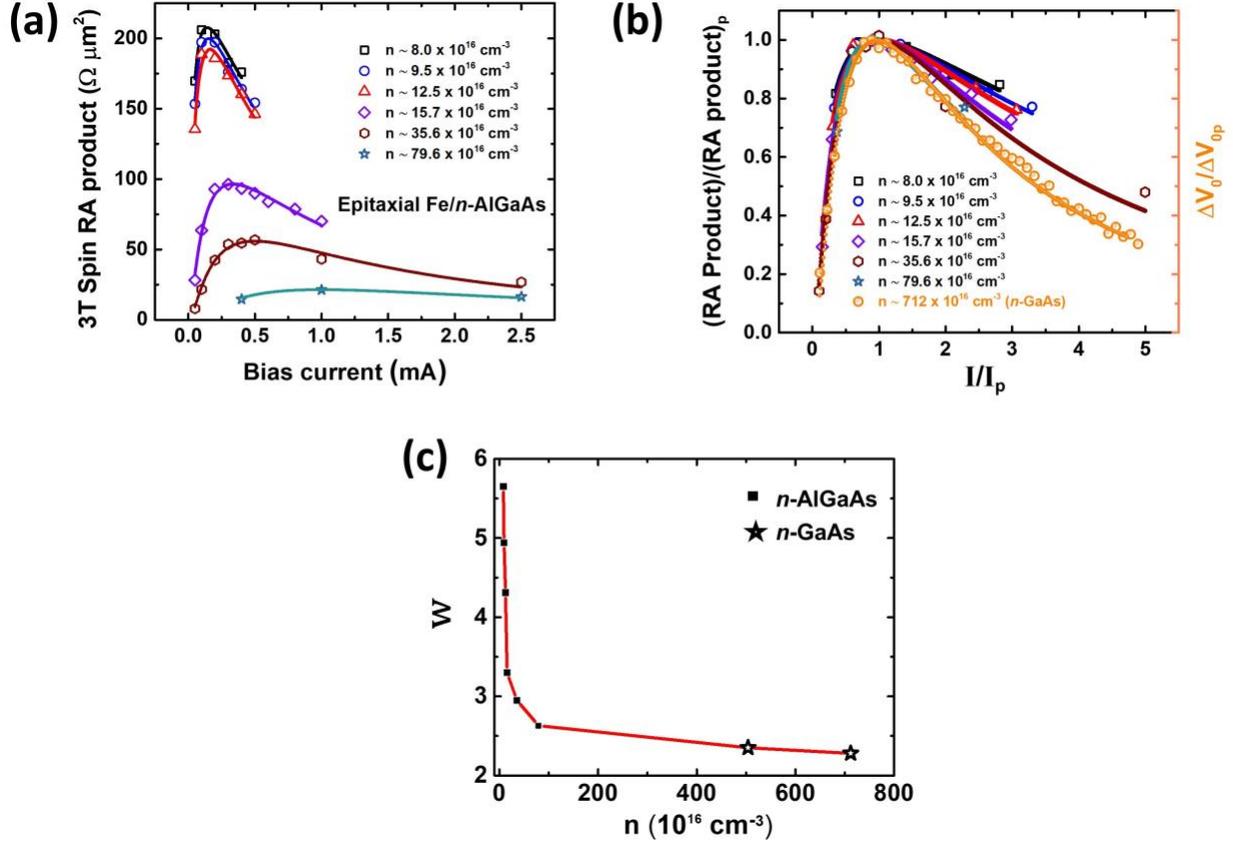

Fig. S10: (a) Bias current dependences of the 3T spin-RA product at varying carrier densities from 3T Hanle measurements of epitaxial Fe/n-AlGaAs heterojunctions [3]. (b) Comparison of the curves in (a) after scaling with the scaled bias-dependence curve of the chiral molecular junctions (n-GaAs) (yellow curve). Here, hollow symbols represent experimental data and solid lines are fits to Eq. 1. (c) Variation of the parameter $w$ with carrier density. Here, the solid squares represent data from the Fe/n-AlGaAs heterojunctions and hollow stars represents data from Au/AHPA-L/n-GaAs junctions.

The behavior of the Hanle signals obtained from chiral molecular junctions exhibits important broad similarities to those in solid-state FM/SC devices. In a previous study of epitaxial FM/SC heterostructures, we took advantage of the persistent photoconductivity of Si-doped AlGaAs (*4*), and obtained Hanle curves at different carrier densities in one and the same Fe/n-AlGaAs devices by photo-doping *via* incremental sub-bandgap illumination (*3*). The bias current dependence of the spin-RA value at varying carrier densities for such a device is shown in Fig. S10(a). The bias dependence of spin RA product is non-monotonic, which is similar to the bias dependence of $\Delta V_0$ in this work. Note that $\Delta V_0$ and the spin-RA product are distinct physical quantities. However, Equation S2 represents a general description of a log-normal behavior, whose width varies with $w$. The normalized curves from Fig. S10(a), along with the one from the chiral molecular junctions on Si-doped GaAs in this study, are shown in Fig. S10(b). The curve narrows with increasing carrier density, which is reflected directly in the smooth decrease of the fitting parameter $w$ with increasing carrier density from fits to Equation S2. A striking aspect here is that the value of $w$ continuously approaches the value 2.3±0.1 in the $n$-GaAs at a carrier density of 7.1x10$^{18}$ cm$^{-3}$ obtained in this work (Fig. S10(c)).

## 10. Determination of Spin Polarization Coefficient

A theoretical model formulated by Valet and Fert (5) predicts a spin accumulation linear with the bias current:

$$\Delta V = \frac{\gamma}{e}(\Delta\mu)_I = \frac{\gamma r_N(\beta r_F + \gamma r_b^*)}{r_F + r_N + r_b^*}j, \tag{S3}$$

where $(\Delta\mu)_I$ is the spin accumulation in the semiconductor channel at the interface, $\gamma$ is the spin-polarization coefficient of the interface, and $\beta$ is the bulk spin-polarization coefficient. $r_F = \rho_F \lambda_{sf}^F$ and $r_N = \rho_N \lambda_{sf}^N$ are the spin resistivity, namely the product of resistivity and spin diffusion length of the FM and SC, respectively. $r_b^*$ is the specific resistivity of the interface between FM and SC, and $j$ is the current density. As $r_F \ll r_N$ and $r_b^*$, Equation S3 is reduced to:

$$\Delta V \approx \frac{\gamma^2 r_N r_b^*}{r_N + r_b^*}j \tag{S4}$$

In this equation, all the parameters are materials dependent. The spin diffusion length ($\lambda_{sf}$) is calculated using diffusion coefficient $D$ and spin lifetime $\tau_s$, $\lambda_{sf} = \sqrt{D\tau_s}$, where for a degenerately doped semiconductor:

$$D = \frac{1}{3}v_f\lambda_s = \frac{1}{3}v_f\left(\frac{\mu m_{eff} v_f}{e}\right) = \frac{1}{3}\left(\frac{\mu m_{eff}}{e}\right)\left[\left(\frac{\hbar}{m_{eff}}\right)(3\pi^2 n)^{1/3}\right]^2.$$

Here, $m_{eff} = 0.067 m_e$ for GaAs [6], $n = 7.1 \times 10^{18}\ cm^{-3}$. As $r_N \ll r_b^*$, Equation S4 is further reduced to:

$$\Delta V \approx \gamma^2 r_N j \tag{S5}$$

By extracting the value of $\Delta V_0$ from the region where it exhibits linearity with respect to current ($I$) (as shown in Fig. 3) and taking the full junction area (5 × 5 μm²) to calculate $j$, we can determine a value of γ, which is found to be 10.0. Table S3 summarizes the values of various parameters:

| $n$ $(cm^{-3})$ | $\rho$ $(m\Omega \cdot cm)$ | $\mu_D$ $(\frac{cm^2}{V \cdot s})$ | $D$ $(m^2/s)$ | $\tau_s$ (ps) | $\lambda_{sf}^N$ (μm) | $r_N$ $(\Omega \cdot m^2)$ | $r_b^*$ $(\Omega \cdot m^2)$ | $\gamma$ |
|---|---|---|---|---|---|---|---|---|
| $7.1 \times 10^{18}$ | 0.52 | 1588 | 0.021 | 273 | 2.4 | $1.24 \times 10^{-11}$ | $1.5 \times 10^{-8}$ | 10.04 |

Table S3: Values of relevant parameters.

The calculated value of γ based on the above equations is unphysical, even though other parameters have reasonable values.

Txoperena *et al.* (*5*) introduced a correction factor to Equation S5 to account for the appropriate current profile at the interface. Considering this correction factor, Equation S5 can be modified to:

$$\Delta V \approx \gamma^2 r_N j * \frac{\lambda_{sf}}{t_{NM}}, \tag{S6}$$

where $t_{NM}$ is the thickness of the NM SC used. In our case, $t_{NM}$ = 0.4 μm. With Equation S6, the value of $\gamma$ is reduced to 4.1, which is still unphysical.

As discussed in detail in the main text, the primary factor contributing to this discrepancy is the value of current density $j$, which is notably small. This observation suggests that the effective current density through the chiral molecule in the junction exceeds the current density value obtained using the entire junction area.

## 11. Power-Law T-dependence from Spin Lifetime

From equation S5, we can infer that $\Delta V \approx \gamma^2 r_N j \propto r_N$, which implies that

$$\Delta V \propto \lambda_{sf}^N \propto \sqrt{\tau_s} \tag{S7}$$

for a degenerately doped semiconductor. Yafet *et al.* (*8,9*) showed that

$$\frac{1}{\tau_s} \propto T^5 \ (6,7). \tag{S8}$$

Combining Equations S7 and S8, one obtains:

$$\Delta V \propto T^{-5/2}. \tag{S9}$$

Such a *T*-dependence contradicts the observations in our experiments.

## 12. Spin Lifetime from Lorentzian Fits

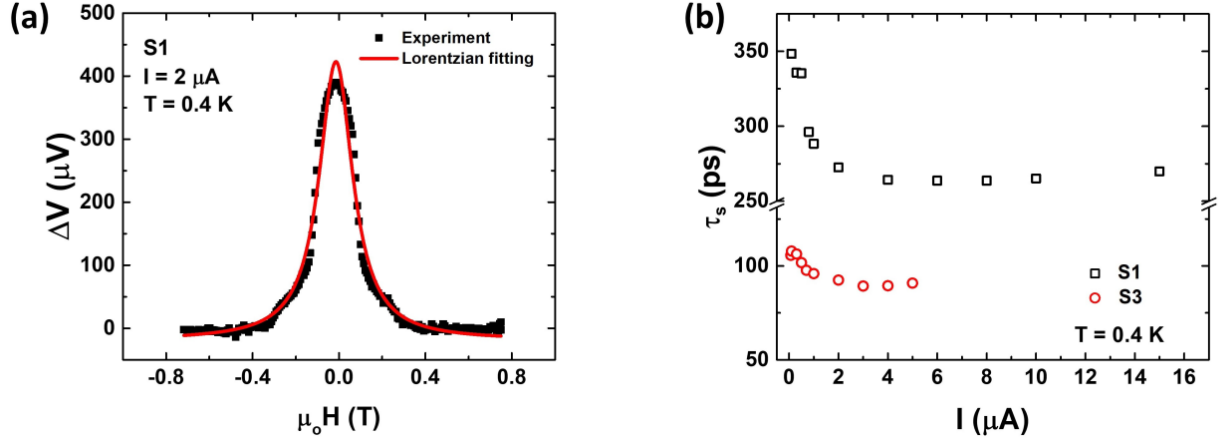

Fig. S11: (a) Lorentzian fits of a Hanle curve in Origin. (b) Bias current dependence of spin lifetime ($\tau_s$) estimated from the full width at half maximum of the Hanle curves for S1 (black square) and S3 (red circle) at 0.4 K.

The spin lifetime ($\tau_s$) can be extracted from Lorentzian fits of the Hanle signals. The Hanle signal can be fit to a Lorentzian function (*10,11*):

$$\Delta V(B) = \frac{\Delta V(0)}{1+\left(\frac{g\mu_B B \tau_s}{\hbar}\right)^2}, \tag{S10}$$

where $\Delta V(0)$ is the magnitude of Hanle signal, $g$ is the electron g-factor, $\mu_B$ is the Bohr magneton, $B$ is the magnetic induction, $\tau_s$ is the spin lifetime, and $\hbar$ is the reduced Planck constant. Here, $\tau_s$ can be calculated from the full width at half maximum (FWHM):

$$\tau_s = \frac{\hbar}{g\mu_B B_{1/2}} = \frac{2\hbar}{g\mu_B W}, \tag{S11}$$

where $W$ = FWHM = $2B_{1/2}$.

In Fig. S11(a), the signal for sample S1 measured at 0.4 K with a bias current of 2 µA is fit to the Lorentzian function. The resulting value of $W$ is then used to calculate $\tau_s$ based on Equation S11. The Hanle curve does not fit well with a well-defined Lorentzian function, however, we can quantitatively extract the FWHM to determine $\tau_s$ approximately. Substituting $g$ with 0.44 (for GaAs (*12*)) and $W$ with 0.189 T, the resulting spin lifetime is 273 ps.

Figure S11(b) shows the $\tau_s$ estimated from the widths of the Hanle curves for all bias currents for S1 and S3 at 0.4 K. The bias dependence of $\tau_s$ is similar in different samples, with the value of $\tau_s$ ranging from 348-270 ps for S1 and 110-90 ps for S3.

## 13. Possible Presence of Dynamic Nuclear Polarization (DNP)

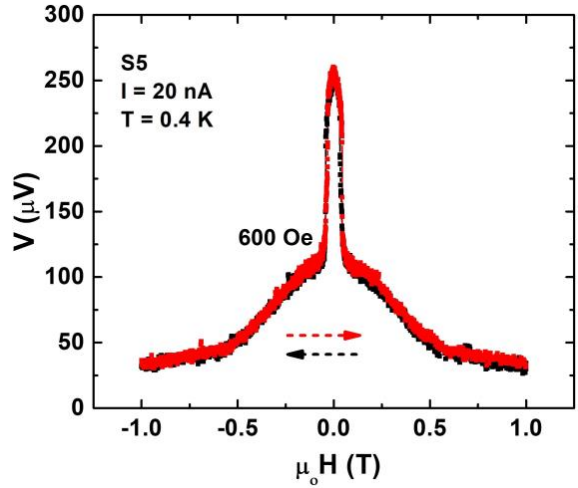

Fig. S12: Hanle signal measured in S5 at 0.4 K with a bias current of 20 nA, showing a sharp peak near zero field on a broad Hanle background.

As discussed in the main text, many of the chiral molecular junctions studied show an apparent superposition of two Lorentzian-like curves. Figure S12 shows the most pronounced example: The Hanle curve from S5 shows two distinct components, with a narrow central peak. This may be due to the effects of dynamic nuclear polarization (DNP). Qualitatively similar features have been observed in conventional *n*-GaAs devices with ferromagnetic injectors (*13*), which was shown to originate from the polarization of nuclear magnetic moments in the GaAs due to hyperfine interactions with injected spin-polarized electrons.